\documentstyle[aps,epsfig]{revtex}

\newcommand{\ket}[1]{%
| #1 \rangle 
}
\newcommand{\matrixel}[3]{%
\langle #1 | #2 | #3 \rangle 
}

\begin{document}

\title{Bosonic behavior of excitons and screening: a consistent calculation}
\author{Christian Tanguy\footnote{phone: 33~1~42~31~78~81; fax: 33~1~42~53~49~30;\\ electronic address: christian.tanguy@rd.francetelecom.com}
}
\address{FRANCE TELECOM R\&D RTA/CDP, 196 avenue Henri Ravera, 92225 Bagneux Cedex, France
}
\date{\today}
\maketitle

\begin{abstract}
Excitons have recently been shown to deviate from pure bosons at densities a hundred times smaller than the Mott density. The corresponding calculations relied on the {\em unscreened} excitonic ground state wavefunction. A consistent inclusion of screening, by use of the fundamental eigenfunction of the Hulth\'{e}n potential, vindicates this approximation.
\vskip0.2cm
\noindent PACS numbers: 71.35Lk
\end{abstract}
\vskip0.3cm


\section{Introduction}

In a recent work\cite{MCCT2001a}, the following question was addressed: Up to what density can excitons be treated as bosons? When the density of electron-hole (e-h) pairs in semiconductors is very large, the carriers form an e-h plasma, so that excitons do not exist anymore. From dimensional arguments, we could expect the cross-over to appear when the distance between two excitons is of the order of their size, i.e., when $\lambda \, N a_x^3/V \simeq 1$, $a_x$ being the exciton Bohr radius, $N$ the number of excitons in the sample volume $V$, and $\lambda$ a dimensionless factor. The high-density regime in which the Coulomb energy is dominated by the kinetic energy is valid when the parameter $r_s$, defined by $N \, \frac{4}{3} \, \pi (r_s \, a_x)^3/V = 1$, is small compared to 1. This leads to $\lambda = 4 \, \pi/3 \simeq 4$. The prefactor $\lambda$ can also be obtained by evaluating the carrier density for which, due to Coulomb screening, no excitonic bound state survives. The accepted result\cite{Klingshirn} for this Mott density leads to $\lambda \simeq 1$.

Still, excitons are {\em not} perfect bosons because they ``feel'' each other not only through Coulomb interaction but also through Pauli exclusion between their electrons and holes\cite{MCRC,Keldysh}. More precisely, the consequences of the fermionic character of excitons induced by Pauli exclusion between their constituents lead to prefactors $\lambda \approx 100$\cite{MCCT2001a}. This final result indicates that our calculations using the three-dimensional fundamental wavefunction of the {\em unscreened} exciton, which admittedly lack proper consistency, should not be too far from the truth, because screening effects are small in that density regime.

In the present Letter, we justify this assumption by considering the fundamental eigenfunction of the Hulth\'{e}n potential\cite{Hulthen,Dow74,Banyai86,HaugKoch,PQM}, which has been widely used to describe three-dimensional screening. The ensuing calculations are analytical throughout, and we basically obtain the same two orders of magnitude for $\lambda$ as previously. If anything, taking account of screening slightly increases $\lambda$.

\section{Notations}

Let us briefly summarize the key quantities introduced in Ref.~\cite{MCCT2001a}. For the sake of simplicity, we will restrict again to excitons with zero total momentum. In terms of the free electron $a^{\dagger}_{k}$ and hole $b^{\dagger}_{-k}$ creation operators, the exact creation operator for the ground-state exciton reads
\begin{equation}
B^{\dagger} = \sum_k \, \phi_k \, a^{\dagger}_{k} \, b^{\dagger}_{-k},
\label{definition B croix}
\end{equation}
where $\phi_k$ is the ground-state exciton wavefunction in momentum space. From Eq.~(\ref{definition B croix}), we get
\begin{equation}
D = 1-[B,B^{\dagger}] = \sum_k \, |\phi_k|^2 \, (a^{\dagger}_{k} \, a_{k} + b^{\dagger}_{-k} \, b_{-k}).
\label{definition D}
\end{equation}
Obviously, $D$ vanishes for perfect bosons.

The relevant state in our problem is the $N$-pair state constructed from the ground-state exciton
\begin{equation}
\ket{\Psi^{(N)}} = \frac{(B^{\dagger})^N \ket{0}}{\sqrt{\matrixel{0}{B^N  \, (B^{\dagger})^N }{0}}} .
\label{definition Psi N excitons}
\end{equation}
This state is the valid picture for the ground state of $N$ pairs in the low-density limit, as shown by Keldysh and Kozlov\cite{KK}. We also showed that the quantity $F_N$ defined by
\begin{equation}
\matrixel{0}{B^N  \, (B^{\dagger})^N}{0} = N! \; F_N
\label{definition F_N}
\end{equation}
plays a central role since it appears in all expectation values taken on $\ket{\Psi^{(N)}}$. If $B$ were a perfect boson operator, we would get $F_N \equiv 1$. Actually, the general expression of $F_N$ reads\cite{MCCT2001a,FN compact}
\begin{equation}
F_N = N! \, \sum_{[i_m]} \hskip3mm \prod_{m=1}^N \, \frac{1}{i_m!} \hskip3mm \bigg( \frac{(-1)^{m+1} \, \sigma_m}{m} \bigg)^{i_m} ,
\label{formule F_N}
\end{equation}
where the $[i_m]$ are determined by $\sum \, m \, i_m = N$, after we have set
\begin{equation}
\sigma_m = \sum_k \, |\phi_k|^{2 m}.
\label{definition sigma_m}
\end{equation}
From the abovementioned expressions we could derive the main result of Ref.~\cite{MCCT2001a}, namely that excitons can be considered as true bosons provided that the expectation value of $D$ defined in Eq.~(\ref{definition D}) is much smaller than unity, which translates into
\begin{equation}
2 \, N \, \sigma_2 \ll 1.
\label{critere D}
\end{equation}

\section{Hulth\'{e}n fundamental wavefunctions}

In contrast to Ref.~\cite{MCCT2001a}, we now want to include some measure of screening in the wavefunction $\phi_k$. This is possible by using the Hulth\'{e}n potential\cite{Hulthen,Dow74,Banyai86,HaugKoch,PQM,CTHulthen}
\begin{equation}
V_H(r) = - \frac{4 \, R}{g} \, \frac{1}{e^{\frac{2 \, r}{g \, a_x}}-1},
\label{V Hulthen}
\end{equation}
where $R$ is the three-dimensional exciton binding energy, $a_x$ its Bohr radius, and $g$ a dimensionless constant inversely proportional to the screening constant derived from the Debye-H\"{u}ckel or Thomas-Fermi models. Only a finite number of bound states remains; they all disappear for $g<1$. $V_H$ has an interesting property: all {\em s} eigenfunctions are known analytically. In the present study however, we are interested in the lowest-lying bound state only. In real space, its eigenfunction reads\cite{HaugKoch}
\begin{equation}
\psi_0({\mathbf r}) = \frac{1}{\sqrt{V}} \, \frac{1}{\sqrt{4 \, \pi}} \, \frac{2}{a_x^{3/2}} \, \sqrt{1-\frac{1}{g^2}} \; \; \; \frac{\sinh(\frac{r}{g \, a_x})}{\frac{r}{g \, a_x}} \; \; \; e^{\displaystyle -\frac{r}{a_x}} .
\label{psi Hulthen r}
\end{equation}
The analytical expression of $\phi_k$ can thus be deduced by a mere Fourier transform of Eq.~(\ref{psi Hulthen r}):
\begin{equation}
\phi_k = \sqrt{\frac{64 \, \pi \, a_x^3}{V} \, (1-\frac{1}{g^2})} \; \; \frac{1}{[(1-\frac{1}{g})^2+K^2] \, [(1+\frac{1}{g})^2+K^2]}, 
\label{phi Hulthen k}
\end{equation}
where $K = k \, a_x$. In the $g \rightarrow +\infty$ limit, we recover the expected result $|\phi_{k}|^2 = z/(1+K^2)^{4}$, where $z = 64 \, \pi \, a_x^3/V$.

\section{Calculation of the new criterion}

By definition, the $\sigma_m$'s given by
\begin{equation}
\sigma_m = \frac{V}{2 \, \pi^2 \, a_x^3} \, \left( \frac{64 \, \pi \, a_x^3}{V} \, \Big(1-\frac{1}{g^2}\Big) \right)^m \, \int_0^{+\infty} \frac{K^2 \, dK}{[(1-\frac{1}{g})^2+K^2]^{2 m} \, [(1+\frac{1}{g})^2+K^2]^{2 m}}
\label{sigma m integrale}
\end{equation}
are now functions of the screening parameter $g$. The first ones are respectively
\begin{eqnarray}
\sigma_1(g) & = & 1\\
\sigma_2(g) & = & \frac{33}{128} \, \tilde{z} \, \left( 1-\frac{2}{3} \, \frac{1}{g^2} + \frac{5}{33} \, \frac{1}{g^4}\right) \label{sigma 2}\\
\sigma_3(g) & = & \frac{4199}{32768} \, \tilde{z}^2 \, \left( 1-\frac{12}{13} \, \frac{1}{g^2} + \frac{6}{13} \, \frac{1}{g^4} -\frac{28}{221} \, \frac{1}{g^6} + \frac{63}{4199} \, \frac{1}{g^8}\right) \label{sigma 3}\\
\sigma_4(g) & = & \frac{334305}{4194304} \, \tilde{z}^3 \, \left( 1-\frac{18}{17} \, \frac{1}{g^2} + \frac{225}{323} \, \frac{1}{g^4} -\frac{100}{323} \, \frac{1}{g^6} + \frac{675}{7249} \, \frac{1}{g^8} -\frac{594}{37145} \, \frac{1}{g^{10}} + \frac{143}{111435} \, \frac{1}{g^{12}}\right) \label{sigma 4}
\end{eqnarray}
with
\begin{equation}
\tilde{z} = \frac{64 \, \pi \, a_x^3}{V \, (1-\frac{1}{g^2})^3}.
\label{z tilde}
\end{equation}
Actually, the general result can be expressed analytically as
\begin{equation}
\sigma_m(g) = (\tilde{z})^{m-1} \; \; \; 16 \, \frac{(8 \, m-5)!!}{(8 \, m-2)!!} \; \; \; {}_2\!F_1\Big(2-2 \, m,\frac{3}{2};2 \, m + \frac{1}{2}; \frac{1}{g^2}\Big),
\label{sigma m analytique}
\end{equation}
where ${}_2\!F_1$ is the hypergeometric function; when $g \rightarrow +\infty$, this function goes to 1, while $\tilde{z}$ goes to $z$, so that we recover the result of Ref.~\cite{MCCT2001a} for the unscreened exciton.

Note that $g$ appears at different places in Eq.~(\ref{sigma m analytique}). The first effect of screening amounts to a ``renormalization'' of the excitonic Bohr radius from $a_x$ to $a_x/(1-1/g^2)$. Because of the decreased electron-hole interaction, the exciton binding energy should decrease and the Bohr radius should therefore increase, tending to infinity at the dissociation limit. This is an unsurprising behavior. None the less, $g$ also appears in the argument of the hypergeometric function, so that the simple physical picture of a dilated exciton is not strictly true.

From Eq.~(\ref{critere D}), we finally obtain for our perfect boson-exciton criterion
\begin{equation}
2 \, N \, \sigma_2(g) = 33 \, \pi \; \; \frac{\displaystyle 1-\frac{2}{3} \, \frac{1}{g^2} + \frac{5}{33} \, \frac{1}{g^4}}{(1-1/g^2)^3} \; \; \frac{N}{V} \, a_x^3 \ll 1 .
\label{condition sigma 2}
\end{equation}

\section{Discussion}

\begin{figure}[htb]
\vskip1.0cm
\hskip4.2cm
\includegraphics[scale=0.3]{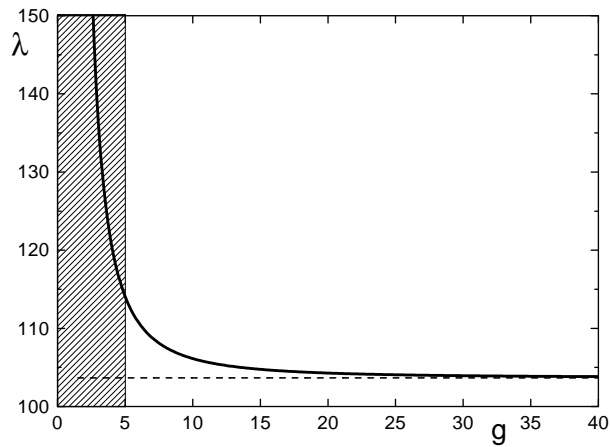}
\caption{Variation of $\lambda$ with the screening parameter $g$.}
\label{lambda(g)}
\end{figure}

Let us now discuss how our consistent introduction of screening affects the criterion for a perfect boson behavior of excitons. We have $\lambda$ equal to $33 \, \pi \, [1-2/(3 \, g^2)  + 5/(33 \, g^4)] \, (1-1/g^2)^{-3}$; its dependence with $g$ is displayed on Fig.~\ref{lambda(g)}. We can see that $\lambda$ increases monotonously as $g$ decreases, and that the change is rather steep only for $g \leq 5$ (hatched region). Otherwise, $\lambda$ stays very close to its asymptotic limit $33 \, \pi \approx 100$.

Next, we have to relate $g$ with the density $n = N/V$. We do so by using an intermediate physical parameter, namely the screening length $\kappa$ given by $\displaystyle \kappa^2 = 4 \, \pi \, e^2 \, \frac{\partial n}{\partial \mu}$ ($\mu$ being the chemical potential). In the case of the Debye-H\"{u}ckel screening or the Thomas-Fermi model at $T=0$, $\kappa^2$ is exactly proportional to $n$\cite{HaugKoch}. In the latter case, at $T \neq 0$, this remains approximately true provided that $n$ is not too large. The second step is obtained by following B\'{a}nyai and Koch to state that $\kappa \, a_x \propto 1/g$\cite{Banyai86}. The value of the proportionality constant depends on how one chooses to correlate the Yukawa and Hulth\'{e}n potentials, but it is always very close to 1\cite{kappa et 1/g?}. From these results, we propose the approximate relationship
\begin{equation}
\frac{1}{g^2} \approx \frac{n}{n_c},
\label{evaluation de g}
\end{equation}
where $n_c$ is the Mott density, thereby ensuring that no bound state survives at $n=n_c$. For densities $n < 10^{-2} \, n_c$, we get $g > 10$. Consequently, the influence of screening on the criterion obtained in Ref.~\cite{MCCT2001a} is hardly significant. If anything, it would lower further the ``critical'' density above which excitons cannot be considered as perfect bosons.

\section{Conclusion}

We have shown that screening may be included consistently in calculations pertaining to the departure of excitons from purely bosonic behavior. This improvement does not alter substantially a former evaluation, in which screening was utterly neglected: Excitons cannot be considered as perfect bosons anymore well below the Mott density (about two orders of magnitude). This should not be without consequence on the possible observation of the quite challenging Bose condensation of excitons.

\section*{Acknowledgements}

Very useful discussions with Dr. Monique Combescot, who introduced me to the problem of the influence of Pauli exclusion on the bosonic behavior of excitons, are gratefully acknowledged.




\begin{references}
\bibitem{FN compact} A more compact expression of $F_N$ is $\displaystyle F_N = \frac{d^N \, \Xi}{d\eta^N} \bigg|_{\eta=0} $ where $\displaystyle \Xi(\eta) = \prod_{k} \, (1 + \eta \, |\phi_k|^{2})$.
\bibitem{kappa et 1/g?} See the note added in proof at the end of Ref.~\cite{Banyai86}, where the origin of the values 1 and $12/\pi^2$ is discussed.
\end{references}
\end{document}